\documentclass[aps,prl,onecolumn,endfloats,preprint,superscriptaddress]{revtex4-1}
\usepackage{amsmath}
\usepackage{graphicx}
\usepackage{longtable}
\usepackage{hyperref}
\usepackage{amssymb}
\usepackage{dcolumn}
\usepackage{multirow}
\usepackage{array}
\usepackage{mhchem}

\begin{document}

\title{Restoration of the Derivative Discontinuity in Kohn-Sham Density Functional Theory: An Efficient Scheme for Energy Gap Correction} 

\date{\today}

\author{Jeng-Da Chai} 
\email[Author to whom correspondence should be addressed. Electronic mail: ]{jdchai@phys.ntu.edu.tw.} 
\affiliation{Department of Physics, National Taiwan University, Taipei 10617, Taiwan} 
\affiliation{Center for Theoretical Sciences and Center for Quantum Science and Engineering, National Taiwan University, Taipei 10617, Taiwan} 

\author{Po-Ta Chen} 
\affiliation{Department of Physics, National Taiwan University, Taipei 10617, Taiwan} 

\begin{abstract} 

From the perspective of perturbation theory, we propose a systematic procedure for the evaluation of the derivative discontinuity (DD) of the exchange-correlation energy functional in Kohn-Sham 
density functional theory (KS-DFT), wherein the exact DD can in principle be obtained by summing up all the perturbation corrections to infinite order. Truncation of the perturbation series at low order 
yields an efficient scheme for obtaining the approximate DD. While the zeroth-order theory yields a vanishing DD, the first-order correction to the DD can be expressed as an explicit universal functional of the 
ground-state density and the KS lowest unoccupied molecular orbital density, allowing the direct evaluation of the DD in the standard KS method without extra computational cost. The fundamental gap 
can be predicted by adding the estimated DD to the KS gap. This scheme is shown to be accurate in the prediction of the fundamental gaps for a wide variety of atoms and molecules. 

\end{abstract}

\maketitle

Over the past two decades, Kohn-Sham density functional theory (KS-DFT) \cite{KSDFT} has become one of the most powerful theoretical methods for studying the ground-state properties of electronic systems. As the 
exact exchange-correlation (XC) energy functional $E_{xc}[\rho]$ in KS-DFT remains unknown, functionals based on the local density approximation (LDA) and generalized gradient approximations (GGAs), have been 
widely used for large systems, due to their computational efficiency and reasonable accuracy. However, owing to their qualitative failures in a number of situations \cite{OD,SciYang,DFTreview,Chai2012}, resolving these 
failures at a reasonable computational cost continues being the subject of intense research interest. 

The prediction of the fundamental gap $E_{g}$ has been an important and challenging subject in 
KS-DFT \cite{G0W0,DD4,GKS1,HSE,RPA,DDcorr,errsinfuncs,DD7,MBJ,GKS2,Hirao,Correction1,HybBG,Correction2,FG115,Correction3}. 
For a system of $N$ electrons ($N$ is an integer) in the presence of an external potential $v_{ext}(\textbf{r})$, $E_{g}$ is defined as 
\begin{equation}\label{eq:FBG}
E_{g} = \text{I}(N) - \text{A}(N), 
\end{equation}
where $\text{I}(N) = E(N-1) - E(N)$ is the vertical ionization potential and $\text{A}(N) = E(N) - E(N+1)$ is the vertical electron affinity, with ${E}({N})$ being the ground-state energy of the $N$-electron system. 
Therefore, $E_{g}$ can be extracted from three KS calculations for the ground-state energies of the $N$- and ($N\pm$1)-electron systems. However, such multiple energy-difference calculations are inapplicable for 
the prediction of fundamental band gaps of solid-state systems \cite{G0W0,DD4,GKS1,RPA,HSE,DDcorr,MBJ,Correction1,HybBG}. 

By contrast, the KS gap ${\Delta}_{\text{KS}}$ is defined as the energy difference between the highest occupied molecular orbital (HOMO) and the lowest unoccupied molecular orbital (LUMO) of the $N$-electron 
system \cite{Janak,Fractional,DD1}, 
\begin{equation}
{\Delta}_{\text{KS}} = {\epsilon}_{N+1}(N) - {\epsilon}_{N}(N), 
\label{eq:KSG}
\end{equation}
where ${\epsilon}_{i}(N)$ is the $i$-th KS orbital energy of the $N$-electron system. Therefore, ${\Delta}_{\text{KS}}$ can be obtained from only one KS calculation for the KS orbital energies of the $N$-electron system. 
Note that $E_{g}$ is not simply ${\Delta}_{\text{KS}}$, but is given by 
\begin{equation}\label{eq:FBG3}
E_{g} = {\Delta}_{\text{KS}} + {\Delta}_{xc}, 
\end{equation}
where 
\begin{equation}\label{eq:DD}
{\Delta}_{xc} = \lim_{\eta \to 0^{+}} \bigg\lbrace \frac{\delta E_{xc}[\rho]}{\delta \rho({\bf r})}\bigg |_{N+\eta} - \frac{\delta E_{xc}[\rho]}{\delta \rho({\bf r})}\bigg |_{N-\eta} \bigg\rbrace 
\end{equation}
is the derivative discontinuity (DD) of $E_{xc}[\rho]$ \cite{Fractional,DD1,DD2,DD2a,DD3,HOMO2,DD4a,DD0,DD6,DD8}. As the KS gap (even with the exact functional) severely underestimates the fundamental 
gap \cite{DD1,DD2,DD6}, the evaluation of the DD is tremendously important. Recently, the importance of the DD in the excited-state \cite{CT} and time-dependent \cite{TD1,TD2} properties has also been highlighted. 
Although several schemes have been proposed for calculating the DD, they can be computationally very demanding for large systems, due to the use of the nonlocal energy-dependent self-energy 
operators \cite{G0W0,DD4,RPA,DDcorr,HybBG} or Hartree-Fock operator \cite{GKS1,DD7,Correction3}. 

In this Letter, we provide a systematic procedure for the evaluation of the DD, based on perturbation theory \cite{Griffiths}. The lowest-order estimate of the DD can be expressed as an explicit 
universal (i.e., system-independent) functional of the ground-state density and the KS LUMO density, allowing very efficient and accurate calculations of the DD and, via Eq.\ (\ref{eq:FBG3}), the fundamental gap 
in the standard KS method. 

For the exact KS-DFT, $\text{I}(N) = - {\epsilon}_{N}(N)$ \cite{Janak,Fractional,Levy84,1overR,HOMO,HOMO2} and therefore $\text{A}(N) = \text{I}(N+1) = - {\epsilon}_{N+1}(N+1)$. 
Consequently, $E_{g}$ (see Eq.\ (\ref{eq:FBG})) can be expressed as 
\begin{equation}\label{eq:FBG2} 
E_{g} = {\epsilon}_{N+1}(N+1) - {\epsilon}_{N}(N), 
\end{equation} 
which is simply the energy difference between the HOMOs of the $N$- and ($N+$1)-electron systems \cite{DD2a}. 
By subtracting Eq.\ (\ref{eq:KSG}) from (\ref{eq:FBG2}), ${\Delta}_{xc}$ (see Eq.\ (\ref{eq:FBG3})) can be expressed as 
\begin{equation}
\begin{split}
\Delta_{xc} =&\; {\epsilon}_{N+1}(N+1) - {\epsilon}_{N+1}(N) \\
=&\; \langle \tilde{\psi}_{N+1} \mid \tilde{H}_{\text{KS}} \mid \tilde{\psi}_{N+1} \rangle - \langle \psi_{N+1} \mid H_{\text{KS}} \mid \psi_{N+1} \rangle. 
\end{split}
\label{eq:d14}
\end{equation} 
Here ${H}_{\text{KS}} \equiv \lbrace -\frac{\hbar^2}{2 m_e} {\bf \nabla}^2 + v_{ext}(\textbf{r}) + e^2 \int \frac{\rho({\bf r'})}{|{\bf r} - {\bf r'}|}d{\bf r'} + v_{xc}([\rho];\textbf{r}) \rbrace$ and $\psi_{i}(\textbf{r})$ are, respectively, 
the KS Hamiltonian and the $i$-th KS orbital of the $N$-electron system, with $v_{xc}([\rho];\textbf{r})$ being the XC potential and $\rho({\bf r}) = \sum_{i=1}^{N} |\psi_{i}({\bf r})|^{2}$ being the ground-state density. 
$\tilde{H}_{\text{KS}} \equiv \lbrace -\frac{\hbar^2}{2 m_e} {\bf \nabla}^2 + v_{ext}(\textbf{r}) + e^2 \int \frac{\tilde{\rho}({\bf r'})}{|{\bf r} - {\bf r'}|}d{\bf r'} + v_{xc}([\tilde{\rho}];\textbf{r}) \rbrace$ and $\tilde{\psi}_{i}(\textbf{r})$ are, 
respectively, the KS Hamiltonian and the $i$-th KS orbital of the ($N$+1)-electron system, with $v_{xc}([\tilde{\rho}];\textbf{r})$ being the XC potential and $\tilde{\rho}({\bf r}) = \sum_{i=1}^{N+1} |\tilde{\psi}_{i}({\bf r})|^{2}$ 
being the ground-state density. 

Aiming to compute ${\Delta}_{xc}$ (and hence $E_{g}$) using only one KS calculation for the $N$-electron system (e.g., for being applicable to solids), we express ${\epsilon}_{N+1}(N+1)$ in terms of 
$\{{\epsilon}_{i}(N), \psi_{i}(\textbf{r})\}$, based on perturbation theory \cite{Griffiths}. 

We choose $H_{\text{KS}}$ as the unperturbed Hamiltonian, and suppose that the unperturbed energy levels are nondegenerate. 
Let $\lambda$ be a dimensionless parameter, ranging continuously from 0 (no perturbation) to 1 (the full perturbation). Consider the perturbed Hamiltonian $H_{\lambda}$ given by 
\begin{equation}\label{eq:s1b}
H_{\lambda} = H_{\text{KS}} + \lambda H^{'}_{\lambda}, 
\end{equation}
where the perturbation $H^{'}_{\lambda} \equiv e^2 \int \frac{ \tilde{\rho}_{\lambda}({\bf r'})}{ | \textbf{ r } - \textbf{ r }^{'} | } d \textbf{r}^{'} + v_{xc}([\tilde{\rho}_{\lambda}];\textbf{r}) 
- e^2 \int \frac{ \rho(\textbf{r}^{'}) } { | \textbf{ r } - \textbf{ r }^{'} | } d \textbf{r}^{'} - v_{xc}([\rho];\textbf{r})$ involves 
$\tilde{\rho}_{\lambda}(\textbf{r}) \equiv \sum^{N+1}_{i=1} | \tilde{\psi}^{\lambda}_{i}(\textbf{r}) |^2$ (filling the orbitals in order of increasing energy). 
Here $\{\tilde{\psi}^{\lambda}_{i}(\textbf{r})\}$ and $\{{\epsilon}^{\lambda}_{i}(N+1)\}$ are, respectively, the eigenstates and eigenvalues of $H_{\lambda}$: 
\begin{equation}
H_{\lambda} \tilde{\psi}^{\lambda}_{i}(\textbf{r}) = {\epsilon}^{\lambda}_{i}(N+1) \tilde{\psi}^{\lambda}_{i}(\textbf{r}). 
\label{eq:s2}
\end{equation} 
Eq.\ (\ref{eq:s2}) at $\lambda = 1$ is simply the KS equation for the ($N$+1)-electron system, as it can be verified that $\{\tilde{\psi}_{i}(\textbf{r})\}$ and $\{{\epsilon}_{i}(N+1)\}$ 
are, respectively, the eigenstates and eigenvalues of $H_{\lambda = 1}$. Therefore, ${\epsilon}_{N+1}(N+1) = {\epsilon}^{\lambda = 1}_{N+1}(N+1)$. 

Writing $H^{'}_{\lambda}$, $\tilde{\psi}^{\lambda}_{i}(\textbf{r})$, and ${\epsilon}^{\lambda}_{i}(N+1)$ as power series in $\lambda$, we have
\begin{equation}\label{eq:s1c}
H^{'}_{\lambda} = H^{'(0)} + {\lambda} H^{'(1)} + {\lambda}^2 H^{'(2)} + \cdots, 
\end{equation}
\begin{equation}\label{eq:s1d}
\tilde{\psi}^{\lambda}_{i}(\textbf{r}) = \psi^{(0)}_{i}(\textbf{r}) + \lambda \psi^{(1)}_{i}(\textbf{r}) + \lambda^2 \psi^{(2)}_{i}(\textbf{r}) + \cdots,
\end{equation} 
\begin{equation}\label{eq:s1e}
{\epsilon}^{\lambda}_{i}(N+1) = \epsilon^{(0)}_{i} + \lambda \epsilon^{(1)}_{i} + {\lambda}^2 \epsilon^{(2)}_{i} + \cdots. 
\end{equation}
Inserting Eqs.\ (\ref{eq:s1b}), (\ref{eq:s1c}), (\ref{eq:s1d}), and (\ref{eq:s1e}) into Eq.\ (\ref{eq:s2}) gives 
\begin{equation}\label{eq:pt}
\begin{split}
&(H_{\text{KS}} + \lambda H^{'(0)} + {\lambda}^2 H^{'(1)} + \cdots)(\psi^{(0)}_{i} + \lambda \psi^{(1)}_{i} + {\lambda}^2 \psi^{(2)}_{i} + \cdots)\\
&= (\epsilon^{(0)}_{i} + \lambda \epsilon^{(1)}_{i} + {\lambda}^2 \epsilon^{(2)}_{i} + \cdots)(\psi^{(0)}_{i} + \lambda \psi^{(1)}_{i} + {\lambda}^2 \psi^{(2)}_{i} + \cdots). 
\end{split}
\end{equation}
Expanding Eq.\ (\ref{eq:pt}) and comparing coefficients of each power of $\lambda$ yields an infinite series of simultaneous equations. 

To zeroth order $(\lambda^0)$ in Eq.\ (\ref{eq:pt}), the equation is 
\begin{equation}
H_{\text{KS}}\psi^{(0)}_{i}(\textbf{r}) = \epsilon^{(0)}_{i} \psi^{(0)}_{i}(\textbf{r}), 
\end{equation}
which is simply the KS equation for the $N$-electron system (i.e., the unperturbed system). 
We then have $\psi^{(0)}_{i}(\textbf{r})$ = $\psi_{i}(\textbf{r})$ and $\epsilon^{(0)}_{i}$ = ${\epsilon}_{i}(N)$. 
Therefore, ${\epsilon}_{N+1}(N+1) = {\epsilon}^{\lambda = 1}_{N+1}(N+1) \approx \epsilon^{(0)}_{N+1} = {\epsilon}_{N+1}(N)$. 
Correspondingly, $\Delta_{xc} = {\epsilon}_{N+1}(N+1) - {\epsilon}_{N+1}(N) \approx {\epsilon}_{N+1}(N) - {\epsilon}_{N+1}(N) = 0$, 
and $E_{g} = {\Delta}_{\text{KS}} + \Delta_{xc} \approx {\Delta}_{\text{KS}}$. Therefore, to obtain a nonvanishing $\Delta_{xc}$, it is necessary to go beyond the zeroth-order theory. 

To first order $(\lambda^1)$ in Eq.\ (\ref{eq:pt}) (see Supplemental Material \cite{Sup}), the first-order correction to the orbital energy is 
\begin{equation}\label{eq:p1b}
\epsilon^{(1)}_{i} = \langle \psi^{(0)}_{i} \mid H^{'(0)} \mid \psi^{(0)}_{i}\rangle = \langle \psi_{i} \mid H^{'}_{\lambda = 0} \mid \psi_{i}\rangle, 
\end{equation}
and the first-order correction to the orbital is 
\begin{equation}\label{eq:p1h1}
\psi^{(1)}_{i}(\textbf{r}) = \sum_{j \neq i} \frac{\langle \psi^{(0)}_{j} \mid H^{'(0)} \mid \psi^{(0)}_{i}\rangle}{\epsilon^{(0)}_{i} - \epsilon^{(0)}_{j}}\psi^{(0)}_{j}(\textbf{r}). 
\end{equation}
Note that $\tilde{\rho}_{\lambda = 0}(\textbf{r})$ = $\sum^{N+1}_{i=1} | \tilde{\psi}^{\lambda = 0}_{i}(\textbf{r}) |^2$ = $\sum^{N+1}_{i=1} |\psi^{(0)}_{i}(\textbf{r})|^2$ = $\sum^{N+1}_{i=1} |\psi_{i}(\textbf{r})|^2$ 
= $\rho(\textbf{r}) + \rho_{\text{L}}(\textbf{r})$, where $\rho_{\text{L}}(\textbf{r}) \equiv |\psi_{N+1}(\textbf{r})|^2$ is the KS LUMO density of the $N$-electron system. 
Consequently, we have 
\begin{equation}
H^{'}_{\lambda = 0} = e^2 \int \frac{\rho_{\text{L}}({\bf r'})}{|\textbf{ r } - \textbf{ r }^{'}|} d \textbf{r}^{'} + v_{xc}([\rho + \rho_{\text{L}}];\textbf{r}) - v_{xc}([\rho];\textbf{r}). 
\end{equation} 
As ${\epsilon}_{N+1}(N+1) = {\epsilon}^{\lambda = 1}_{N+1}(N+1) \approx \epsilon^{(0)}_{N+1} + \epsilon^{(1)}_{N+1}$, we have 
\begin{equation}\label{eq:p1j}
\begin{split}
\Delta_{xc} &= {\epsilon}_{N+1}(N+1) - {\epsilon}_{N+1}(N) \\ 
& \approx \epsilon^{(1)}_{N+1} = \langle \psi_{N+1} \mid H^{'}_{\lambda = 0} \mid \psi_{N+1}\rangle \\
&= e^2 \iint \frac{\rho_{\text{L}}({\bf r})\rho_{\text{L}}({\bf r'})}{|{\bf r} - {\bf r'}|}d{\bf r}d{\bf r'} \\
&\;\;\; + \int \rho_{\text{L}}(\textbf{r}) \bigg\lbrace v_{xc}([\rho + \rho_{\text{L}}];\textbf{r}) - v_{xc}([\rho];\textbf{r}) \bigg\rbrace d\textbf{r}, 
\end{split}
\end{equation}
and $E_{g} = {\Delta}_{\text{KS}} + \Delta_{xc} \approx {\Delta}_{\text{KS}} + \epsilon^{(1)}_{N+1}$. 
Eq.\ (\ref{eq:p1j}) is a key result, showing that the DD can be approximately expressed as an explicit universal functional of $\rho({\bf r})$ and $\rho_{\text{L}}({\bf r})$, and 
can be calculated in the standard KS method without extra computational cost. 
Note that it can also be derived from Eq.\ (\ref{eq:d14}) by assuming (``frozen orbital approximation (FOA)") that 
\begin{equation}
\tilde{\psi}_{i}(\textbf{r}) \approx \psi_{i}(\textbf{r}),\; i = 1, 2, 3, \cdots. 
\label{eq:new}
\end{equation}  

To second order $(\lambda^2)$ in Eq.\ (\ref{eq:pt}) (see Supplemental Material \cite{Sup}), the second-order correction to the orbital energy is 
\begin{equation}\label{eq:p2b}
\begin{split}
\epsilon^{(2)}_{i} &= \langle \psi^{(0)}_{i} \mid H^{'(0)} \mid \psi^{(1)}_{i}\rangle + \langle \psi^{(0)}_{i} \mid H^{'(1)} \mid \psi^{(0)}_{i}\rangle\\
&= \sum_{j \neq i} \frac{|\langle \psi_{j} \mid H^{'}_{\lambda = 0} \mid \psi_{i}\rangle|^2}{{\epsilon}_{i}(N) - {\epsilon}_{j}(N)} + \langle \psi_{i} \mid H^{'(1)} \mid \psi_{i}\rangle, 
\end{split}
\end{equation}
where 
\begin{equation}
\begin{split}
H^{'(1)} &= \frac{\partial H^{'}_{\lambda}}{\partial \lambda}{\bigg|}_{\lambda = 0} \\
&= \int \bigg\lbrace \frac{e^2} { | \textbf{ r } - \textbf{ r }^{'} | } + \frac{\delta v_{xc}([\tilde{\rho}_{\lambda}];\textbf{r})} {\delta \tilde{\rho}_{\lambda}({\bf r}')} \bigg\rbrace 
\frac{\partial \tilde{\rho}_{\lambda}(\textbf{r}^{'})}{\partial \lambda} {\bigg|}_{\lambda = 0} d \textbf{r}^{'} \\ 
&= \int \bigg\lbrace \frac{e^2} { | \textbf{ r } - \textbf{ r }^{'} | } + f_{xc}([\rho + \rho_{\text{L}}];\textbf{r},\textbf{r}') \bigg\rbrace \\ 
&\;\;\; \times \bigg\lbrace 2 \sum_{i = 1}^{N+1} {\Re}(\psi^{*}_{i}(\textbf{r}') \psi^{(1)}_{i}(\textbf{r}')) \bigg\rbrace d \textbf{r}^{'}. 
\end{split}
\end{equation} 
Here $f_{xc}([\rho];\textbf{r},\textbf{r}') \equiv \delta v_{xc}([\rho];\textbf{r}) / \delta \rho({\bf r}')$ is the XC kernel, the asterisk (*) denotes complex conjugate, and ${\Re}(\cdots)$ denotes 
the real part of $(\cdots)$. From Eq.\ (\ref{eq:p2b}), we have 
\begin{equation}\label{eq:p2b1}
\epsilon^{(2)}_{N+1} = \sum_{j \neq N+1} \frac{|\langle \psi_{j} \mid H^{'}_{\lambda = 0} \mid \psi_{N+1}\rangle|^2}{{\epsilon}_{N+1}(N) - {\epsilon}_{j}(N)} + \langle \psi_{N+1} \mid H^{'(1)} \mid \psi_{N+1}\rangle. 
\end{equation}
Correspondingly, ${\epsilon}_{N+1}(N+1) = {\epsilon}^{\lambda = 1}_{N+1}(N+1) \approx \epsilon^{(0)}_{N+1} + \epsilon^{(1)}_{N+1} + \epsilon^{(2)}_{N+1}$. 
This gives $\Delta_{xc} = {\epsilon}_{N+1}(N+1) - {\epsilon}_{N+1}(N) \approx \epsilon^{(1)}_{N+1} + \epsilon^{(2)}_{N+1}$ 
and $E_{g} = {\Delta}_{\text{KS}} + \Delta_{xc} \approx {\Delta}_{\text{KS}} + (\epsilon^{(1)}_{N+1} + \epsilon^{(2)}_{N+1})$. 

Extending the process further, the HOMO energy of the ($N$+1)-electron system can be obtained by summing up all the perturbation corrections to infinite order, i.e., 
${\epsilon}_{N+1}(N+1) = {\epsilon}^{\lambda = 1}_{N+1}(N+1) = \sum_{n = 0}^{\infty} \epsilon^{(n)}_{N+1}$. Therefore, we can, in principle, obtain the exact 
$\Delta_{xc} = {\epsilon}_{N+1}(N+1) - {\epsilon}_{N+1}(N) = \sum_{n = 1}^{\infty} \epsilon^{(n)}_{N+1}$ 
and the exact $E_{g} = {\Delta}_{\text{KS}} + \Delta_{xc} = {\Delta}_{\text{KS}} + \sum_{n = 1}^{\infty} \epsilon^{(n)}_{N+1}$. 

For any finite order truncation of the above perturbation series, if two or more unperturbed states share the same energy, degenerate perturbation theory may be needed \cite{Griffiths}. 
Since the concept of the perturbation to the unperturbed Hamiltonian ${H}_{\text{KS}}$ remains valid, 
this scheme could be extended to estimate the $\Delta_{xc}$ (and hence the $E_{g}$) for the degenerate cases based on the corresponding degenerate perturbation theory. 

As mentioned previously, the DD needs to be summed to the KS gap to give the fundamental gap. While the DD given by Eq.\ (\ref{eq:DD}) should be the same as that given by Eq.\ (\ref{eq:d14}) for the exact functional, 
this property may {\it no longer} hold true for an approximate functional. For example, for a LDA or GGA, while the DD given by Eq.\ (\ref{eq:DD}) is shown to vanish \cite{DD0,DD7,Correction3}, we emphasize that 
the DD can be {\it favorably restored} by Eq.\ (\ref{eq:d14}), and subsequently approximated by Eq.\ (\ref{eq:p1j}). Although a more accurate approximation for the DD could be pursued by higher-order perturbation theory, 
we adopt the DD given by Eq.\ (\ref{eq:p1j}) (i.e., first-order correction) for simplicity. 
Accordingly, the fundamental gap is predicted by summing Eqs.\ (\ref{eq:KSG}) and (\ref{eq:p1j}) in our ${\Delta}_{\text{KS}} + {\Delta}_{xc}$ scheme. 

Here we examine the performance of various schemes in the prediction of the fundamental gaps for the FG115 database \cite{FG115}, which consists of 115 accurate reference values for the fundamental gaps of 
18 atoms and 97 molecules at their experimental geometries. The fundamental gaps are calculated by our ${\Delta}_{\text{KS}} + {\Delta}_{xc}$ scheme, the ${\Delta}_{\text{KS}}$ scheme (by Eq.\ (\ref{eq:KSG})), 
and the $E_{g}$ scheme (by Eq.\ (\ref{eq:FBG2})), using the LDA \cite{LDA} and LB94 \cite{LB94} functionals, and the 6-311++G(3df,3pd) basis set, 
with a development version of \textsf{Q-Chem 3.2} \cite{QChem}. The error for each entry is defined as (error = theoretical value $-$ reference value). The notation used for characterizing statistical errors is as follows: 
mean signed errors (MSEs), mean absolute errors (MAEs), and root-mean-square (rms) errors. Note that for the ${\Delta}_{\text{KS}}$ or ${\Delta}_{\text{KS}} + {\Delta}_{xc}$ scheme, only one KS calculation 
for the $N$-electron system is required (i.e., applicable to solids), while for the $E_{g}$ scheme, which is the ${\Delta}_{\text{KS}} + {\Delta}_{xc}$ scheme with ${\Delta}_{xc}$ being exactly calculated 
by Eq.\ (\ref{eq:d14}) (with no further approximations), two KS calculations for the $N$- and ($N+$1)-electron systems are required (i.e., inapplicable to solids). 

\begin{figure}[htbp]
\includegraphics[scale=1.0]{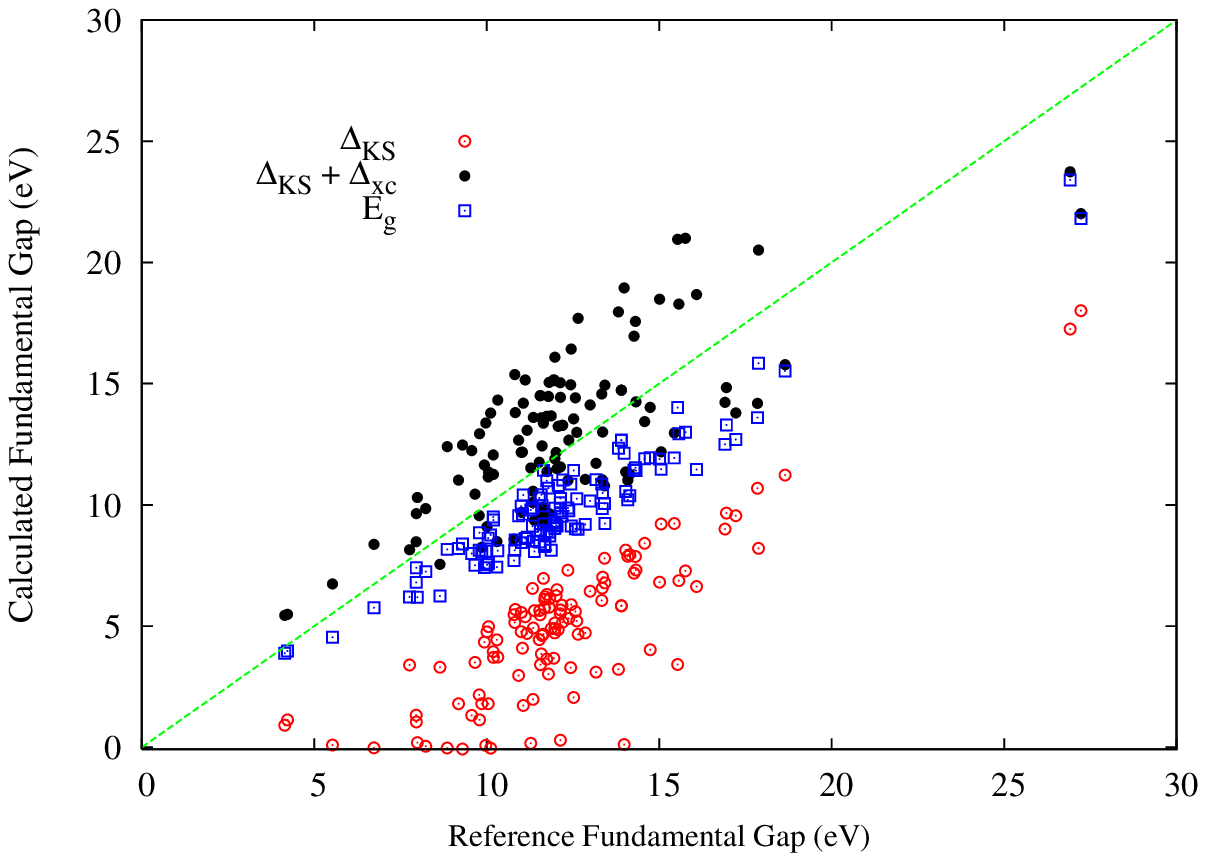}
\caption{\label{f1} Calculated versus reference fundamental gaps for the FG115 database \cite{FG115}. The fundamental gaps are calculated by three schemes (see text for details) using the LDA functional.} 
\end{figure}

\begin{figure}[htbp]
\includegraphics[scale=1.0]{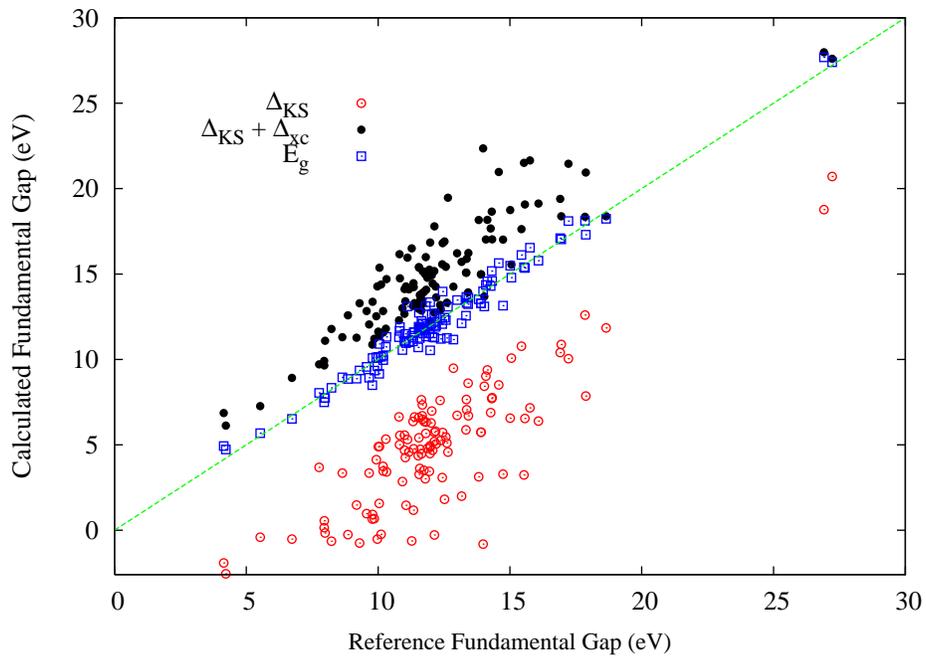}
\caption{\label{f2} Same as Fig.\ \ref{f1}, but using the LB94 functional.} 
\end{figure}

The calculated gaps are plotted against the reference values in Fig.\ \ref{f1} (for LDA) and Fig.\ \ref{f2} (for LB94). For both functionals, as the ${\Delta}_{\text{KS}}$ gaps are shown to be vanishingly small (some of them 
are even negative) for the small-gap (smaller than 10 eV) systems, the DDs are essential for the accurate prediction of the fundamental gaps. In fact, even for the large-gap (larger than 15 eV) systems, the DDs remain 
significant fractions of the fundamental gaps. As shown in Table \ref{t1}, the MAE associated with the ${\Delta}_{\text{KS}} + {\Delta}_{xc}$ or $E_{g}$ scheme is more than three times smaller than 
that associated with the ${\Delta}_{\text{KS}}$ scheme \cite{Sup}. 

Due to the use of ``FOA" (see Eq.\ (\ref{eq:new})) in the evaluation of the DD, the ${\Delta}_{\text{KS}} + {\Delta}_{xc}$ gaps tend to be larger than the $E_{g}$ gaps. 
For rare gas atoms (e.g., He, Ne, and Ar), where this approximation becomes excellent, the ${\Delta}_{\text{KS}} + {\Delta}_{xc}$ gaps are very close to the $E_{g}$ gaps \cite{Sup}. 

For LB94, the $E_{g}$ gaps are in excellent agreement with the reference values, due to the correct asymptote of the LB94 potential, which is a key factor for the accurate prediction of the HOMO 
energies \cite{HOMO2,Levy84,1overR,LB94} and, via Eq.\ (\ref{eq:FBG2}), the fundamental gaps. By contrast, for LDA, the $E_{g}$ gaps tend to underestimate the reference values, due to the imbalanced self-interaction 
errors (as the LDA potential is asymptotically incorrect) in the predicted HOMO energies of the $N$- and ($N$+1)-electron systems \cite{OD,DFTreview,HOMO2,Levy84,1overR,LB94}. As the 
${\Delta}_{\text{KS}} + {\Delta}_{xc}$ gaps tend to overestimate the $E_{g}$ gaps, it appears that there is a fortuitous cancellation of errors in the predicted ${\Delta}_{\text{KS}} + {\Delta}_{xc}$ gaps, when compared with 
the reference values. 

In conclusion, we have provided a systematic procedure for the direct evaluation of the DD, based on perturbation theory. The lowest-order estimate of the DD is an explicit universal functional of 
the ground-state density and the KS LUMO density (see Eq.\ (\ref{eq:p1j})), presenting a simple, efficient, and nonempirical scheme for the direct evaluation of the DD in the standard KS method. 
The fundamental gap can be accurately predicted by the sum of the KS gap and the estimated DD. 
The validity and accuracy of this scheme has been demonstrated for a wide variety of atoms and molecules, extending the applicability of KS-DFT to an area long believed to be beyond its reach. 
To further improve the accuracy of this scheme, a more accurate functional and a more accurate approximation for the DD (based on higher-order perturbation theory) should be adopted, although this will necessarily be 
somewhat more expensive. Since the concepts of the DD and the perturbation to the unperturbed Hamiltonian ${H}_{\text{KS}}$ are still valid for solid-state systems, this scheme 
could be extended to estimate the DD (a correction to the KS band gap) for solids, where the prediction of accurate fundamental band gaps is very challenging for KS-DFT. Work in this direction is in progress. 

This work was supported by National Science Council of Taiwan (Grant No. NSC101-2112-M-002-017-MY3), National Taiwan University (Grant Nos.\ 99R70304, 101R891401, and 101R891403), 
and National Center for Theoretical Sciences of Taiwan. We would like to thank Prof. Shih-I Chu for providing us Refs.\ \cite{TD1,TD2}. 

\bibliographystyle{prl}

\newpage

\newpage

\newpage
\begin{table*}
\footnotesize
\caption{\label{t1} Statistical errors (in eV) of the 115 fundamental gaps of the FG115 database \cite{FG115}, calculated by three schemes (see text for details) using the LDA and LB94 functionals.} 
\begin{ruledtabular} 
\begin{tabular}{lcccccc} 
& \multicolumn{2}{r}{LDA\;\;\;\;\;} & \multicolumn{3}{r}{LB94\;\;\;\;\;}\\
\cline{2-4}
\cline{5-7}
Error & ${\Delta}_{\text{KS}}$ & ${\Delta}_{\text{KS}}+{\Delta}_{xc}$ & $E_{g}$ & ${\Delta}_{\text{KS}}$ & ${\Delta}_{\text{KS}}+{\Delta}_{xc}$ & $E_{g}$ \\ \hline 
MSE & -7.22 & 0.78 & -2.33 & -7.20 & 2.73 &  0.03 \\ 
MAE &  7.22 & 2.11 &   2.33 &  7.20 & 2.74 &  0.47 \\ 
rms   &  7.44 & 2.45 &   2.57 &  7.48 & 3.13 &  0.63 \\ 
\end{tabular}
\end{ruledtabular}
\end{table*}

\newpage
\section*{Supplemental Material to: Restoration of the Derivative Discontinuity in Kohn-Sham Density Functional Theory: An Efficient Scheme for Energy Gap Correction}

\section{Derivative discontinuity from the perspective of \\ perturbation theory}

As discussed in our paper, we aim to express ${\epsilon}_{N+1}(N+1)$ in terms of $\{{\epsilon}_{i}(N), \psi_{i}(\textbf{r})\}$, based on perturbation theory \cite{Griffiths}. 
Here, we provide the full derivation (the notations used here are the same as those used in the paper). As Eq.\ (12) of our paper gives 
\begin{equation}\label{eq:pt}
\begin{split}
&(H_{\text{KS}} + \lambda H^{'(0)} + {\lambda}^2 H^{'(1)} + \cdots)(\psi^{(0)}_{i} + \lambda \psi^{(1)}_{i} + {\lambda}^2 \psi^{(2)}_{i} + \cdots)\\
&= (\epsilon^{(0)}_{i} + \lambda \epsilon^{(1)}_{i} + {\lambda}^2 \epsilon^{(2)}_{i} + \cdots)(\psi^{(0)}_{i} + \lambda \psi^{(1)}_{i} + {\lambda}^2 \psi^{(2)}_{i} + \cdots), 
\end{split}
\end{equation}
expanding Eq.\ (\ref{eq:pt}) and comparing coefficients of each power of $\lambda$ yields an infinite series of simultaneous equations.

\subsection{Zeroth-Order Theory} 

To zeroth order $(\lambda^0)$ in Eq.\ (\ref{eq:pt}), the equation is 
\begin{equation}
H_{\text{KS}}\psi^{(0)}_{i}(\textbf{r}) = \epsilon^{(0)}_{i} \psi^{(0)}_{i}(\textbf{r}), 
\end{equation}
which is simply the KS equation for the $N$-electron system (i.e., the unperturbed system). 
We then have $\psi^{(0)}_{i}(\textbf{r})$ = $\psi_{i}(\textbf{r})$ and $\epsilon^{(0)}_{i}$ = ${\epsilon}_{i}(N)$. 
Therefore, ${\epsilon}_{N+1}(N+1) = {\epsilon}^{\lambda = 1}_{N+1}(N+1) \approx \epsilon^{(0)}_{N+1} = {\epsilon}_{N+1}(N)$. 
Correspondingly, $\Delta_{xc} = {\epsilon}_{N+1}(N+1) - {\epsilon}_{N+1}(N) \approx {\epsilon}_{N+1}(N) - {\epsilon}_{N+1}(N) = 0$, 
and $E_{g} = {\Delta}_{\text{KS}} + \Delta_{xc} \approx {\Delta}_{\text{KS}}$. Therefore, to obtain a nonvanishing $\Delta_{xc}$, it is necessary to go beyond the zeroth-order theory.

\subsection{First-Order Theory} 

To first order $(\lambda^1)$ in Eq.\ (\ref{eq:pt}), the equation is 
\begin{equation}\label{eq:p1}
H_{\text{KS}}\psi^{(1)}_{i}(\textbf{r}) + H^{'(0)}\psi^{(0)}_{i}(\textbf{r}) = \epsilon^{(0)}_{i} \psi^{(1)}_{i}(\textbf{r}) + \epsilon^{(1)}_{i} \psi^{(0)}_{i}(\textbf{r}).
\end{equation}
Taking the inner product of Eq.\ (\ref{eq:p1}) with $\psi^{(0)}_{i}$, 
\begin{equation}\label{eq:p1a}
\begin{split}
&\langle \psi^{(0)}_{i} \mid H_{\text{KS}} \mid \psi^{(1)}_{i}\rangle + \langle \psi^{(0)}_{i} \mid H^{'(0)} \mid \psi^{(0)}_{i}\rangle \\
&=\epsilon^{(0)}_{i} \langle \psi^{(0)}_{i}\mid \psi^{(1)}_{i}\rangle + \epsilon^{(1)}_{i} \langle \psi^{(0)}_{i} \mid \psi^{(0)}_{i}\rangle.
\end{split}
\end{equation}
As $H_{\text{KS}}$ is Hermitian, the first term on the left-hand side is the same as that on the right-hand side. Moreover, we have 
$\langle \psi^{(0)}_{i} \mid \psi^{(0)}_{i}\rangle = \langle \psi_{i} \mid \psi_{i}\rangle = 1$, so the first-order correction to the orbital energy is 
\begin{equation}\label{eq:p1b}
\epsilon^{(1)}_{i} = \langle \psi^{(0)}_{i} \mid H^{'(0)} \mid \psi^{(0)}_{i}\rangle = \langle \psi_{i} \mid H^{'}_{\lambda = 0} \mid \psi_{i}\rangle.
\end{equation}
Rewriting Eq.\ (\ref{eq:p1}), we have 
\begin{equation}\label{eq:p1c}
(H_{\text{KS}} - \epsilon^{(0)}_{i})\psi^{(1)}_{i}(\textbf{r}) = -(H^{'(0)} - \epsilon^{(1)}_{i})\psi^{(0)}_{i}(\textbf{r}).
\end{equation}
Since $\{\psi^{(0)}_{j}(\textbf{r})\}$ constitute a complete set, $\psi^{(1)}_{i}(\textbf{r})$ can be expressed as a linear combination of them: 
\begin{equation}\label{eq:p1d}
\psi^{(1)}_{i}(\textbf{r}) = \sum_{j \neq i} a^{(i)}_{j} \psi^{(0)}_{j}(\textbf{r}). 
\end{equation}
Note that it is unnecessary to include $j = i$ in the sum, for if $\psi^{(1)}_{i}(\textbf{r})$ satisfies Eq.\ (\ref{eq:p1c}), so too does $(\psi^{(1)}_{i}(\textbf{r}) + \alpha \psi^{(0)}_{i}(\textbf{r}))$, for any constant $\alpha$, 
and we can use this freedom to subtract off the $\psi^{(0)}_{i}(\textbf{r})$ term so that $\tilde{\psi}^{\lambda}_{i}(\textbf{r})$ is normalized at first order in $\lambda$. 
Inserting Eq.\ (\ref{eq:p1d}) into Eq.\ (\ref{eq:p1c}), we have 
\begin{equation}\label{eq:p1e}
\sum_{j \neq i} (\epsilon^{(0)}_{j} - \epsilon^{(0)}_{i})a^{(i)}_{j}\psi^{(0)}_{j}(\textbf{r}) = -(H^{'(0)} - \epsilon^{(1)}_{i})\psi^{(0)}_{i}(\textbf{r}). 
\end{equation}
Taking the inner product of Eq.\ (\ref{eq:p1e}) with $\psi^{(0)}_{k}$, we get 
\begin{equation}\label{eq:p1f}
\begin{split}
&\sum_{j \neq i} (\epsilon^{(0)}_{j} - \epsilon^{(0)}_{i})a^{(i)}_{j} \langle \psi^{(0)}_{k} \mid \psi^{(0)}_{j}\rangle \\
&= - \langle \psi^{(0)}_{k} \mid H^{'(0)} \mid \psi^{(0)}_{i}\rangle + \epsilon^{(1)}_{i}  \langle \psi^{(0)}_{k} \mid \psi^{(0)}_{i}\rangle.
\end{split}
\end{equation}
If $k = i$, the left-hand side is zero, we recover Eq.\ (\ref{eq:p1b}); If $k \ne i$, we get $(\epsilon^{(0)}_{k} - \epsilon^{(0)}_{i}) a^{(i)}_{k} = - \langle \psi^{(0)}_{k} \mid H^{'(0)} \mid \psi^{(0)}_{i}\rangle$, 
or $a^{(i)}_{j} = \frac{\langle \psi^{(0)}_{j} \mid H^{'(0)} \mid \psi^{(0)}_{i}\rangle}{\epsilon^{(0)}_{i} - \epsilon^{(0)}_{j}}$, so the first-order correction to the orbital is 
\begin{equation}\label{eq:p1h1}
\psi^{(1)}_{i}(\textbf{r}) = \sum_{j \neq i} \frac{\langle \psi^{(0)}_{j} \mid H^{'(0)} \mid \psi^{(0)}_{i}\rangle}{\epsilon^{(0)}_{i} - \epsilon^{(0)}_{j}}\psi^{(0)}_{j}(\textbf{r}). 
\end{equation}
Note that $\tilde{\rho}_{\lambda = 0}(\textbf{r})$ = $\sum^{N+1}_{i=1} | \tilde{\psi}^{\lambda = 0}_{i}(\textbf{r}) |^2$ = $\sum^{N+1}_{i=1} |\psi^{(0)}_{i}(\textbf{r})|^2$ = $\sum^{N+1}_{i=1} |\psi_{i}(\textbf{r})|^2$ 
= $\rho(\textbf{r}) + \rho_{\text{L}}(\textbf{r})$, where $\rho_{\text{L}}(\textbf{r}) \equiv |\psi_{N+1}(\textbf{r})|^2$ is the KS LUMO density of the $N$-electron system. 
Consequently, we have 
\begin{equation}
H^{'}_{\lambda = 0} = e^2 \int \frac{\rho_{\text{L}}({\bf r'})}{|\textbf{ r } - \textbf{ r }^{'}|} d \textbf{r}^{'} + v_{xc}([\rho + \rho_{\text{L}}];\textbf{r}) - v_{xc}([\rho];\textbf{r}). 
\end{equation} 
As ${\epsilon}_{N+1}(N+1) = {\epsilon}^{\lambda = 1}_{N+1}(N+1) \approx \epsilon^{(0)}_{N+1} + \epsilon^{(1)}_{N+1}$, we have 
\begin{equation}\label{eq:p1j}
\begin{split}
\Delta_{xc} &= {\epsilon}_{N+1}(N+1) - {\epsilon}_{N+1}(N) \\ 
& \approx \epsilon^{(1)}_{N+1} = \langle \psi_{N+1} \mid H^{'}_{\lambda = 0} \mid \psi_{N+1}\rangle \\
&= e^2 \iint \frac{\rho_{\text{L}}({\bf r})\rho_{\text{L}}({\bf r'})}{|{\bf r} - {\bf r'}|}d{\bf r}d{\bf r'} \\
&\;\;\; + \int \rho_{\text{L}}(\textbf{r}) \bigg\lbrace v_{xc}([\rho + \rho_{\text{L}}];\textbf{r}) - v_{xc}([\rho];\textbf{r}) \bigg\rbrace d\textbf{r}, 
\end{split}
\end{equation}
and $E_{g} = {\Delta}_{\text{KS}} + \Delta_{xc} \approx {\Delta}_{\text{KS}} + \epsilon^{(1)}_{N+1}$. 
Eq.\ (\ref{eq:p1j}) is a key result, showing that the DD can be approximately expressed as an explicit universal functional of $\rho({\bf r})$ and $\rho_{\text{L}}({\bf r})$, and 
can be calculated in the standard KS method without extra computational cost.

\subsection{Second-Order Theory}

To second order $(\lambda^2)$ in Eq.\ (\ref{eq:pt}), the equation is 
\begin{equation}\label{eq:p2}
\begin{split}
&H_{\text{KS}}\psi^{(2)}_{i}(\textbf{r}) + H^{'(0)}\psi^{(1)}_{i}(\textbf{r}) + H^{'(1)}\psi^{(0)}_{i}(\textbf{r}) \\
&= \epsilon^{(0)}_{i} \psi^{(2)}_{i}(\textbf{r}) + \epsilon^{(1)}_{i} \psi^{(1)}_{i}(\textbf{r}) + \epsilon^{(2)}_{i} \psi^{(0)}_{i}(\textbf{r}). 
\end{split}
\end{equation}
Taking the inner product of Eq.\ (\ref{eq:p2}) with $\psi^{(0)}_{i}$, we get 
\begin{equation}\label{eq:p2a}
\begin{split}
&\langle \psi^{(0)}_{i} \mid H_{\text{KS}} \mid \psi^{(2)}_{i}\rangle + \langle \psi^{(0)}_{i} \mid H^{'(0)} \mid \psi^{(1)}_{i}\rangle + \langle \psi^{(0)}_{i} \mid H^{'(1)} \mid \psi^{(0)}_{i}\rangle \\ 
&= \epsilon^{(0)}_{i} \langle \psi^{(0)}_{i} \mid \psi^{(2)}_{i}\rangle + \epsilon^{(1)}_{i} \langle \psi^{(0)}_{i} \mid \psi^{(1)}_{i}\rangle + \epsilon^{(2)}_{i} \langle \psi^{(0)}_{i} \mid \psi^{(0)}_{i}\rangle. 
\end{split}
\end{equation}
As $H_{\text{KS}}$ is Hermitian, the first term on the left-hand side is the same as that on the right-hand side. Moreover, we have $\langle \psi^{(0)}_{i} \mid \psi^{(0)}_{i}\rangle = 1$ and 
$\langle \psi^{(0)}_{i} \mid \psi^{(1)}_{i}\rangle = 0$ (see Eq.\ (\ref{eq:p1h1})), so the second-order correction to the orbital energy is 
\begin{equation}\label{eq:p2b}
\begin{split}
\epsilon^{(2)}_{i} &= \langle \psi^{(0)}_{i} \mid H^{'(0)} \mid \psi^{(1)}_{i}\rangle + \langle \psi^{(0)}_{i} \mid H^{'(1)} \mid \psi^{(0)}_{i}\rangle\\
&= \sum_{j \neq i} \frac{|\langle \psi_{j} \mid H^{'}_{\lambda = 0} \mid \psi_{i}\rangle|^2}{{\epsilon}_{i}(N) - {\epsilon}_{j}(N)} + \langle \psi_{i} \mid H^{'(1)} \mid \psi_{i}\rangle, 
\end{split}
\end{equation}
where 
\begin{equation}
\begin{split}
H^{'(1)} &= \frac{\partial H^{'}_{\lambda}}{\partial \lambda}{\bigg|}_{\lambda = 0} \\
&= \int \bigg\lbrace \frac{e^2} { | \textbf{ r } - \textbf{ r }^{'} | } + \frac{\delta v_{xc}([\tilde{\rho}_{\lambda}];\textbf{r})} {\delta \tilde{\rho}_{\lambda}({\bf r}')} \bigg\rbrace 
\frac{\partial \tilde{\rho}_{\lambda}(\textbf{r}^{'})}{\partial \lambda} {\bigg|}_{\lambda = 0} d \textbf{r}^{'} \\ 
&= \int \bigg\lbrace \frac{e^2} { | \textbf{ r } - \textbf{ r }^{'} | } + f_{xc}([\rho + \rho_{\text{L}}];\textbf{r},\textbf{r}') \bigg\rbrace \\ 
&\;\;\; \times \bigg\lbrace 2 \sum_{i = 1}^{N+1} {\Re}(\psi^{*}_{i}(\textbf{r}') \psi^{(1)}_{i}(\textbf{r}')) \bigg\rbrace d \textbf{r}^{'}. 
\end{split}
\end{equation} 
Here $f_{xc}([\rho];\textbf{r},\textbf{r}') \equiv \delta v_{xc}([\rho];\textbf{r}) / \delta \rho({\bf r}')$ is the XC kernel, the asterisk (*) denotes complex conjugate, and ${\Re}(\cdots)$ denotes 
the real part of $(\cdots)$. From Eq.\ (\ref{eq:p2b}), we have 
\begin{equation}\label{eq:p2b1}
\epsilon^{(2)}_{N+1} = \sum_{j \neq N+1} \frac{|\langle \psi_{j} \mid H^{'}_{\lambda = 0} \mid \psi_{N+1}\rangle|^2}{{\epsilon}_{N+1}(N) - {\epsilon}_{j}(N)} + \langle \psi_{N+1} \mid H^{'(1)} \mid \psi_{N+1}\rangle. 
\end{equation}
Correspondingly, ${\epsilon}_{N+1}(N+1) = {\epsilon}^{\lambda = 1}_{N+1}(N+1) \approx \epsilon^{(0)}_{N+1} + \epsilon^{(1)}_{N+1} + \epsilon^{(2)}_{N+1}$. 
This gives $\Delta_{xc} = {\epsilon}_{N+1}(N+1) - {\epsilon}_{N+1}(N) \approx \epsilon^{(1)}_{N+1} + \epsilon^{(2)}_{N+1}$ 
and $E_{g} = {\Delta}_{\text{KS}} + \Delta_{xc} \approx {\Delta}_{\text{KS}} + (\epsilon^{(1)}_{N+1} + \epsilon^{(2)}_{N+1})$.

\newpage
\begin{table}
\footnotesize
\caption{Fundamental gaps (in eV) of the FG115 database \cite{FG115}, calculated by three schemes (see text for details) using the LDA \cite{LDA} and LB94 \cite{LB94} functionals.} 
\begin{ruledtabular}
\begin{tabular}{lccccccc} 
& \multicolumn{3}{r}{LDA\;\;\;\;\;} & \multicolumn{3}{r}{LB94\;\;\;\;\;}\\
\cline{3-5}
\cline{6-8}
Molecule & Reference & ${\Delta}_{\text{KS}}$ & ${\Delta}_{\text{KS}}+{\Delta}_{xc}$ & $E_{g}$ & ${\Delta}_{\text{KS}}$ & ${\Delta}_{\text{KS}}+{\Delta}_{xc}$ & $E_{g}$ \\ \hline 
H (Hydrogen atom) & 12.86 & 4.71 & 11.05 & 9.19 & 9.48 & 14.26 & 11.16\\
He (Helium atom) & 27.23 & 18.01 & 22.01 & 21.81 & 20.71 & 27.60 & 27.40\\
Li (Lithium atom) & 4.22 & 1.13 & 5.49 & 3.98 & -2.55 & 6.13 & 4.72\\
Be (Beryllium atom) & 9.66 & 3.50 & 10.44 & 7.51 & 3.34 & 12.05 & 8.93\\
B (Boron atom) & 7.99 & 0.19 & 10.30 & 6.19 & -0.16 & 11.09 & 7.74\\
C (Carbon atom) & 9.97 & 0.08 & 13.38 & 8.08 & -0.52 & 14.28 & 9.59\\
N (Nitrogen atom) & 14.74 & 4.02 & 14.02 & 11.94 & 3.28 & 17.01 & 13.15\\
O (Oxygen atom) & 12.14 & 0.29 & 14.44 & 10.27 & -0.28 & 17.79 & 12.20\\
F (Fluorine atom) & 13.98 & 0.12 & 18.95 & 12.13 & -0.81 & 22.36 & 13.99\\
Ne (Neon atom) & 26.91 & 17.25 & 23.74 & 23.40 & 18.77 & 27.98 & 27.69\\
Na (Sodium atom) & 4.14 & 0.91 & 5.44 & 3.88 & -1.92 & 6.86 & 4.93\\
Mg (Magnesium atom) & 7.76 & 3.39 & 8.15 & 6.20 & 3.68 & 9.71 & 8.04\\
Al (Aluminum atom) & 5.53 & 0.09 & 6.74 & 4.54 & -0.41 & 7.26 & 5.68\\
Si (Silicon atom) & 6.73 & -0.02 & 8.37 & 5.76 & -0.52 & 8.91 & 6.52\\
P (Phosphorus atom) & 9.78 & 2.16 & 9.56 & 8.13 & 0.65 & 10.87 & 8.49\\
S (Sulfur atom) & 8.23 & 0.04 & 9.85 & 7.25 & -0.64 & 11.79 & 8.34\\
Cl (Chlorine atom) & 9.30 & -0.07 & 12.47 & 8.38 & -0.76 & 13.29 & 9.35\\
Ar (Argon atom) & 18.65 & 11.23 & 15.78 & 15.52 & 11.85 & 18.38 & 18.23\\
\ce{ CH3 } (Methyl radical) & 9.86 & 1.79 & 8.25 & 8.04 & 0.67 & 11.21 & 9.35\\
\ce{ CH4 } (Methane) & 15.06 & 9.21 & 12.19 & 11.47 & 10.08 & 15.55 & 14.78\\
\ce{ NH } (Imidogen) ($^{3}{\Sigma}^{-}$) & 13.17 & 3.10 & 11.72 & 11.04 & 1.99 & 15.71 & 12.11\\
\ce{ NH2 } (Amino radical) & 11.34 & 1.97 & 10.56 & 9.87 & 1.17 & 14.46 & 11.27\\
\ce{ NH3 } (Ammonia) & 11.54 & 5.64 & 9.42 & 8.47 & 6.59 & 12.96 & 11.91\\
\ce{ OH } (Hydroxyl radical) & 11.27 & 0.16 & 11.52 & 9.79 & -0.63 & 16.50 & 11.38\\
\ce{ H2O } (Water) & 13.35 & 6.57 & 11.03 & 9.85 & 7.65 & 15.08 & 13.66\\
\ce{ HF } (Hydrogen fluoride) & 16.91 & 9.00 & 14.23 & 12.50 & 10.40 & 19.39 & 17.10\\
\ce{ SiH3 } (Silyl) & 7.95 & 1.32 & 8.48 & 6.80 & 0.15 & 9.91 & 7.49\\
\ce{ SiH4 } (Silane) & 14.03 & 8.14 & 11.36 & 10.55 & 8.43 & 13.69 & 13.11\\
\ce{ PH3 } (Phosphine) & 11.82 & 6.13 & 9.63 & 8.72 & 6.32 & 14.08 & 11.34\\
\ce{ H2S } (Hydrogen sulfide) & 11.00 & 5.57 & 9.69 & 8.44 & 5.56 & 14.12 & 11.06\\
\ce{ HCl } (Hydrogen sulfide) & 13.36 & 7.02 & 13.00 & 10.49 & 7.06 & 15.89 & 13.30\\
\ce{ C2H2 } (Acetylene) & 13.43 & 6.78 & 14.94 & 9.23 & 6.70 & 16.23 & 13.61\\
\ce{ C2H4 } (Ethylene) & 12.57 & 5.60 & 14.41 & 9.01 & 5.46 & 15.42 & 12.30\\
\ce{ C2H6 } (Ethane) & 13.41 & 7.80 & 10.79 & 10.06 & 8.61 & 13.93 & 13.24\\
\ce{ HCN } (Hydrogen cyanide) & 14.31 & 7.87 & 17.57 & 11.53 & 7.71 & 18.65 & 15.17\\
\ce{ CO } (Carbon monoxide) & 15.57 & 6.87 & 18.28 & 12.93 & 6.54 & 19.07 & 15.35\\
\ce{ HCO } (Formyl radical) & 9.56 & 1.31 & 12.24 & 7.99 & 0.97 & 12.83 & 9.56\\
\ce{ H2CO } (Formaldehyde) & 11.56 & 3.41 & 14.51 & 8.94 & 3.27 & 15.37 & 11.86\\
\ce{ CH3OH } (Methyl alcohol) & 11.67 & 5.75 & 9.76 & 8.28 & 6.70 & 13.87 & 12.07\\
\ce{ N2 } (Nitrogen diatomic) & 17.88 & 8.21 & 20.51 & 15.84 & 7.85 & 20.94 & 17.30\\
\ce{ N2H4 } (Hydrazine) & 10.29 & 4.43 & 8.49 & 7.44 & 5.33 & 11.80 & 10.74\\
\ce{ NO } (Nitric oxide) & 10.11 & -0.04 & 13.79 & 8.76 & -0.24 & 14.38 & 10.15\\
\ce{ O2 } (Oxygen diatomic) ($^{3}{\Sigma}_{g}$) & 12.52 & 2.06 & 13.55 & 11.42 & 1.81 & 16.91 & 12.44\\
\ce{ H2O2 } (Hydrogen peroxide) & 12.65 & 4.66 & 17.70 & 8.99 & 4.57 & 19.47 & 13.08\\
\ce{ F2 } (Fluorine diatomic) & 15.53 & 3.42 & 20.95 & 14.02 & 3.24 & 21.51 & 15.40\\
\ce{ CO2 } (Carbon dioxide) & 14.58 & 8.41 & 13.44 & 11.90 & 8.51 & 20.97 & 15.64\\
\ce{ P2 } (Phosphorus diatomic) & 10.19 & 3.71 & 11.25 & 9.37 & 3.46 & 11.58 & 9.96\\
\ce{ S2 } (Sulfur diatomic) ($^{3}{\Sigma}_{g}$) & 7.96 & 1.05 & 9.64 & 7.40 & 0.55 & 9.65 & 7.75\\
\ce{ Cl2 } (Chlorine diatomic) & 10.93 & 2.96 & 12.67 & 9.55 & 2.85 & 13.00 & 10.55\\
\ce{ NaCl } (Sodium Chloride) & 8.64 & 3.30 & 7.55 & 6.24 & 3.35 & 11.30 & 8.95\\
\ce{ SiO } (Silicon monoxide) & 11.60 & 4.61 & 12.44 & 9.96 & 4.74 & 13.48 & 11.67\\
\ce{ CS } (Carbon monosulfide) & 11.58 & 3.85 & 13.59 & 10.42 & 3.62 & 13.77 & 11.26\\
\ce{ ClO } (Monochlorine monoxide) & 8.85 & -0.03 & 12.41 & 8.16 & -0.26 & 12.58 & 8.86\\
\ce{ ClF } (Chlorine monofluoride) & 12.43 & 3.28 & 14.96 & 10.86 & 3.08 & 15.57 & 12.07\\
\ce{ Si2H6 } (Disilane) & 11.33 & 6.56 & 10.09 & 9.12 & 6.37 & 13.13 & 11.05\\
\ce{ CH3Cl } (Methyl chloride) & 12.01 & 6.25 & 12.16 & 9.16 & 6.28 & 15.05 & 12.09\\
\ce{ CH3SH } (Methanethiol) & 10.01 & 4.76 & 9.10 & 7.53 & 4.89 & 11.62 & 10.14\\
\ce{ SO2 } (Sulfur dioxide) & 11.74 & 3.64 & 13.66 & 10.99 & 3.51 & 13.95 & 11.78\\
\ce{ BF3 } (Borane, trifluoro-) & 17.22 & 9.55 & 13.80 & 12.70 & 10.04 & 21.46 & 18.10\\
\ce{ BCl3 } (Borane, trichloro-) & 12.07 & 4.85 & 13.24 & 10.80 & 4.68 & 14.44 & 11.31\\
\ce{ AlCl3 } (Aluminum trichloride) & 12.13 & 5.66 & 11.57 & 9.67 & 5.73 & 12.73 & 11.27\\
\ce{ CF4 } (Carbon tetrafluoride) & 17.85 & 10.68 & 14.18 & 13.60 & 12.60 & 18.35 & 18.12\\
\ce{ CCl4 } (Carbon tetrachloride) & 11.97 & 4.90 & 11.90 & 9.86 & 4.86 & 12.23 & 10.54\\
\ce{ OCS } (Carbonyl sulfide) & 12.13 & 5.51 & 15.04 & 10.09 & 5.20 & 15.17 & 12.69\\
\ce{ CS2 } (Carbon disulfide) & 10.19 & 3.94 & 12.06 & 9.50 & 3.74 & 12.83 & 10.21\\
\ce{ CF2O } (Carbonic difluoride) & 16.08 & 6.63 & 18.68 & 11.46 & 6.39 & 19.13 & 15.78\\
\ce{ SiF4 } (Silicon tetrafluoride) & 16.95 & 9.66 & 14.84 & 13.30 & 10.87 & 18.38 & 17.03\\
\ce{ N2O } (Nitrous oxide) & 15.01 & 6.81 & 18.48 & 11.89 & 6.56 & 18.74 & 15.49\\
\ce{ NF3 } (Nitrogen trifluoride) & 15.76 & 7.28 & 21.00 & 13.00 & 7.17 & 21.65 & 16.54\\
\ce{ PF3 } (Phosphorus trifluoride) & 13.00 & 6.43 & 14.13 & 10.16 & 6.72 & 16.22 & 13.48\\
\ce{ O3 } (Ozone) & 11.06 & 1.72 & 14.19 & 10.40 & 1.46 & 14.27 & 10.95\\
\ce{ F2O } (Difluorine monoxide) & 13.82 & 3.22 & 17.96 & 12.34 & 3.13 & 18.16 & 13.52\\
\ce{ ClF3 } (Chlorine trifluoride) & 11.79 & 3.02 & 14.48 & 10.69 & 3.02 & 14.98 & 11.72\\
\ce{ C2F4 } (Tetrafluoroethylene) & 12.45 & 5.88 & 16.43 & 9.13 & 5.72 & 16.80 & 13.97\\
\ce{ CH3CCH } (Propyne) & 11.69 & 6.21 & 9.29 & 8.32 & 6.49 & 15.16 & 11.54\\
\ce{ CH2CCH2 } (Allene) & 10.83 & 5.69 & 13.81 & 8.55 & 5.55 & 14.75 & 11.89\\
\ce{ C3H4 } (Cyclopropene) & 11.87 & 4.90 & 13.67 & 8.13 & 4.79 & 14.78 & 11.94\\
\ce{ C3H6 } (Cyclopropane) & 11.64 & 6.97 & 9.91 & 9.21 & 7.64 & 12.86 & 12.25\\
\ce{ CH2F2 } (Methane, difluoro-) & 14.15 & 7.94 & 11.27 & 10.37 & 9.39 & 18.17 & 14.57\\
\ce{ CHF3 } (Methane, trifluoro-) & 15.44 & 9.23 & 12.97 & 11.94 & 10.78 & 17.62 & 16.11\\
\ce{ CH2Cl2 } (Methylene chloride) & 12.18 & 5.86 & 13.28 & 9.52 & 5.80 & 14.27 & 11.91\\
\ce{ CHCl3 } (Chloroform) & 12.38 & 5.31 & 12.67 & 9.75 & 5.26 & 12.92 & 11.20\\
\ce{ CH3NO2 } (Methane, nitro-) & 11.94 & 3.67 & 15.16 & 9.24 & 3.44 & 15.25 & 11.52\\
\ce{ CH3SiH3 } (Methyl silane) & 12.35 & 7.30 & 11.00 & 9.87 & 7.60 & 13.18 & 12.42\\
\ce{ HCOOH } (Formic acid) & 11.98 & 5.13 & 16.09 & 9.03 & 4.95 & 16.84 & 13.35\\
\ce{ CH3CONH2 } (Acetamide) & 10.05 & 4.98 & 11.35 & 7.59 & 4.89 & 15.37 & 10.91\\
\ce{ C2N2 } (Cyanogen) & 13.90 & 5.83 & 14.73 & 12.67 & 5.74 & 14.98 & 13.28\\
\ce{ CH2CO } (Ketene) &10.32&3.72&14.32&8.11&3.42&14.70 &11.32\\
\ce{ C2H4O } (Ethylene oxide) & 11.68 & 6.10 & 9.16 & 8.37 & 7.34 & 13.25 & 12.10\\
\ce{ C2H2O2 } (Ethanedial) & 10.04 & 1.80 & 11.15 & 8.63 & 1.57 & 11.37 & 9.16\\
\ce{ CH3CH2OH } (Ethanol) & 11.38 & 5.64 & 9.37 & 8.07 & 6.63 & 13.30 & 11.60\\
\ce{ CH3OCH3 } (Dimethyl ether) & 10.79 & 5.47 & 8.59 & 7.71 & 6.64 & 12.29 & 11.31\\
\ce{ C2H4S } (Thiirane) & 9.93 & 4.34 & 11.64 & 7.41 & 4.13 & 12.53 & 10.12\\
\ce{ CH2CHF } (Ethene, fluoro-) & 11.55 & 5.46 & 14.50 & 8.78 & 5.41 & 15.41 & 12.73\\
\ce{ CH3CH2Cl } (Ethyl chloride) & 11.74 & 6.31 & 11.37 & 8.90 & 6.36 & 15.13 & 11.85\\
\ce{ CH2CHCl } (Ethene, chloro-) & 11.35 & 4.92 & 13.61 & 8.50 & 4.76 & 14.23 & 11.59\\
\ce{ CH3COCl } (Acetyl Chloride) & 11.97 & 4.72 & 15.12 & 9.29 & 4.48 & 15.16 & 11.79\\
\ce{ NO2 } (Nitrogen dioxide) & 9.79 & 1.13 & 12.94 & 8.85 & 0.93 & 13.37 & 10.08\\
\ce{ CFCl3 } (Trichloromonofluoromethane) & 12.61 & 5.21 & 12.99 & 10.26 & 5.11 & 13.32 & 11.26\\
\ce{ CF3Cl } (Methane, chlorotrifluoro-) & 14.27 & 7.18 & 16.96 & 11.44 & 6.89 & 17.67 & 14.30\\
\ce{ HCCF } (Fluoroacetylene) & 12.04 & 6.50 & 11.49 & 9.12 & 6.98 & 14.93 & 12.64\\
\ce{ HCCCN } (Cyanoacetylene) & 12.20 & 5.16 & 13.29 & 11.03 & 5.05 & 13.63 & 12.00\\
\ce{ C4N2 } (2-Butynedinitrile) & 11.52 & 4.44 & 11.76 & 10.28 & 4.37 & 11.94 & 10.70\\
\ce{ C2N2 } (Cyanogen) & 13.90 & 5.83 & 14.73 & 12.67 & 5.74 & 14.98 & 13.28\\
\ce{ C3O2 } (Carbon suboxide) & 11.64 & 4.67 & 13.37 & 11.42 & 4.47 & 13.53 & 11.82\\
\ce{ FCN } (Cyanogen fluoride) & 14.33 & 7.31 & 14.25 & 11.41 & 7.76 & 17.03 & 14.59\\
\ce{ C4H2 } (Diacetylene) & 11.00 & 4.77 & 12.17 & 9.94 & 4.69 & 12.67 & 11.09\\
\ce{ H2CS } (Thioformaldehyde) & 9.18 & 1.80 & 11.02 & 8.20 & 1.48 & 11.27 & 8.87\\
\ce{ CHONH2 } (Formamide) & 10.81 & 5.15 & 15.38 & 8.13 & 5.00 & 16.16 & 11.67\\
\ce{ CH2CCl2 } (Ethene, 1,1-dichloro-) & 11.17 & 4.69 & 13.08 & 8.66 & 4.58 & 14.11 & 11.01\\
\ce{ C2HF3 } (Trifluoroethylene) & 11.11 & 5.39 & 15.16 & 8.62 & 5.31 & 15.96 & 13.10\\
\ce{ CH2CF2 } (Ethene, 1,1-difluoro-) & 11.81 & 5.77 & 15.06 & 8.87 & 5.66 & 15.99 & 13.12\\
\ce{ CH3F } (Methyl fluoride) & 14.09 & 7.88 & 11.02 & 10.21 & 9.02 & 17.02 & 14.36\\
\ce{ CF2Cl2 } (Difluorodichloromethane) & 13.33 & 6.06 & 14.58 & 10.86 & 5.87 & 15.08 & 12.56\\
\ce{ SiF2 } (Silicon difluoride) & 11.04 & 4.09 & 12.17 & 9.62 & 4.26 & 13.40 & 11.50\\
\hline
MSE &  & -7.22 & 0.78 & -2.33 & -7.20 & 2.73 &  0.03 \\ 
MAE &  &  7.22 & 2.11 &   2.33 &  7.20 & 2.74 &  0.47 \\
rms   &  &  7.44 & 2.45 &   2.57 &  7.48  & 3.13 &  0.63 \\ 	 	 	
\end{tabular}
\end{ruledtabular}
\end{table}

\begin{table}
\footnotesize
\caption{Comparison of errors (in eV) of the 115 derivative discontinuities (in eV) of the FG115 database \cite{FG115}, calculated by the DD(0th) and DD(1st), using the LDA \cite{LDA} and LB94 \cite{LB94} functionals. 
Here DD(0th) = 0 is obtained from the zeroth-order perturbation theory, and DD(1st), which is calculated by the difference between the ${\Delta}_{\text{KS}} + {\Delta}_{xc}$ and ${\Delta}_{\text{KS}}$ gaps, is obtained from the 
first-order perturbation theory (i.e., ``frozen orbital approximation"). The reference (DD) values are calculated by the difference between the $E_{g}$ and ${\Delta}_{\text{KS}}$ gaps, using the respective functionals. 
Note that since LB94 yields very accurate $E_{g}$, and possibly accurate ${\Delta}_{\text{KS}}$ (due to the correct asymptote of the LB94 potential), the reference DD values calculated by LB94 should be more accurate 
than those calculated by LDA.} 
\begin{ruledtabular}
\begin{tabular}{lcccccc} 
& \multicolumn{3}{r}{LDA\;\;\;\;\;\;\;\;\;\;\;\;\;\;\;\;\;\;\;\;} & \multicolumn{3}{r}{LB94\;\;\;\;\;\;\;\;\;\;\;\;\;\;\;\;\;\;\;\;}\\
\cline{2-4}
\cline{5-7}
Molecule  & DD & DD(0th) & DD(1st) & DD & DD(0th) & DD(1st) \\ \hline 
H (Hydrogen atom) & 4.48 & -4.48 & 1.86 & 1.68 & -1.68 & 3.10 \\
He (Helium atom) & 3.80 & -3.80 & 0.20 & 6.70 & -6.70 & 0.20 \\
Li (Lithium atom) & 2.85 & -2.85 & 1.51 & 7.27 & -7.27 & 1.40 \\
Be (Beryllium atom) & 4.01 & -4.01 & 2.93 & 5.58 & -5.58 & 3.12 \\
B (Boron atom) & 6.00 & -6.00 & 4.12 & 7.90 & -7.90 & 3.36 \\
C (Carbon atom) & 8.01 & -8.01 & 5.29 & 10.11 & -10.11 & 4.69 \\
N (Nitrogen atom) & 7.92 & -7.92 & 2.08 & 9.87 & -9.87 & 3.86 \\
O (Oxygen atom) & 9.98 & -9.98 & 4.17 & 12.48 & -12.48 & 5.59 \\
F (Fluorine atom) & 12.02 & -12.02 & 6.82 & 14.80 & -14.80 & 8.37 \\
Ne (Neon atom) & 6.15 & -6.15 & 0.33 & 8.92 & -8.92 & 0.30 \\
Na (Sodium atom) & 2.97 & -2.97 & 1.56 & 6.85 & -6.85 & 1.93 \\
Mg (Magnesium atom) & 2.81 & -2.81 & 1.95 & 4.36 & -4.36 & 1.67 \\
Al (Aluminum atom) & 4.44 & -4.44 & 2.20 & 6.10 & -6.10 & 1.58 \\
Si (Silicon atom) & 5.78 & -5.78 & 2.61 & 7.03 & -7.03 & 2.40 \\
P (Phosphorus atom) & 5.96 & -5.96 & 1.44 & 7.84 & -7.84 & 2.38 \\
S (Sulfur atom) & 7.20 & -7.20 & 2.60 & 8.98 & -8.98 & 3.45 \\
Cl (Chlorine atom) & 8.46 & -8.46 & 4.09 & 10.11 & -10.11 & 3.93 \\
Ar (Argon atom) & 4.29 & -4.29 & 0.26 & 6.38 & -6.38 & 0.15 \\
\ce{ CH3 } (Methyl radical) & 6.24 & -6.24 & 0.21 & 8.67 & -8.67 & 1.86 \\
\ce{ CH4 } (Methane) & 2.26 & -2.26 & 0.72 & 4.71 & -4.71 & 0.77 \\
\ce{ NH } (Imidogen) ($^{3}{\Sigma}^{-}$) & 7.94 & -7.94 & 0.68 & 10.12 & -10.12 & 3.60 \\
\ce{ NH2 } (Amino radical) & 7.90 & -7.90 & 0.70 & 10.10 & -10.10 & 3.19 \\
\ce{ NH3 } (Ammonia) & 2.83 & -2.83 & 0.95 & 5.32 & -5.32 & 1.05 \\
\ce{ OH } (Hydroxyl radical) & 9.63 & -9.63 & 1.74 & 12.01 & -12.01 & 5.12 \\
\ce{ H2O } (Water) & 3.28 & -3.28 & 1.19 & 6.01 & -6.01 & 1.41 \\
\ce{ HF } (Hydrogen fluoride) & 3.50 & -3.50 & 1.73 & 6.70 & -6.70 & 2.29 \\
\ce{ SiH3 } (Silyl) & 5.48 & -5.48 & 1.68 & 7.34 & -7.34 & 2.42 \\
\ce{ SiH4 } (Silane) & 2.41 & -2.41 & 0.81 & 4.67 & -4.67 & 0.58 \\
\ce{ PH3 } (Phosphine) & 2.59 & -2.59 & 0.92 & 5.01 & -5.01 & 2.74 \\
\ce{ H2S } (Hydrogen sulfide) & 2.88 & -2.88 & 1.25 & 5.50 & -5.50 & 3.06 \\
\ce{ HCl } (Hydrogen sulfide) & 3.47 & -3.47 & 2.51 & 6.25 & -6.25 & 2.59 \\
\ce{ C2H2 } (Acetylene) & 2.45 & -2.45 & 5.71 & 6.91 & -6.91 & 2.62 \\
\ce{ C2H4 } (Ethylene) & 3.41 & -3.41 & 5.41 & 6.84 & -6.84 & 3.12 \\
\ce{ C2H6 } (Ethane) & 2.26 & -2.26 & 0.73 & 4.62 & -4.62 & 0.69 \\
\ce{ HCN } (Hydrogen cyanide) & 3.66 & -3.66 & 6.04 & 7.46 & -7.46 & 3.48 \\
\ce{ CO } (Carbon monoxide) & 6.06 & -6.06 & 5.35 & 8.81 & -8.81 & 3.72 \\
\ce{ HCO } (Formyl radical) & 6.68 & -6.68 & 4.25 & 8.59 & -8.59 & 3.27 \\
\ce{ H2CO } (Formaldehyde) & 5.52 & -5.52 & 5.58 & 8.59 & -8.59 & 3.51 \\
\ce{ CH3OH } (Methyl alcohol) & 2.53 & -2.53 & 1.48 & 5.37 & -5.37 & 1.79 \\
\ce{ N2 } (Nitrogen diatomic) & 7.63 & -7.63 & 4.67 & 9.45 & -9.45 & 3.64 \\
\ce{ N2H4 } (Hydrazine) & 3.00 & -3.00 & 1.05 & 5.41 & -5.41 & 1.06 \\
\ce{ NO } (Nitric oxide) & 8.80 & -8.80 & 5.03 & 10.39 & -10.39 & 4.23 \\
\ce{ O2 } (Oxygen diatomic) ($^{3}{\Sigma}_{g}$) & 9.36 & -9.36 & 2.13 & 10.63 & -10.63 & 4.47 \\
\ce{ H2O2 } (Hydrogen peroxide) & 4.33 & -4.33 & 8.71 & 8.51 & -8.51 & 6.39 \\
\ce{ F2 } (Fluorine diatomic) & 10.60 & -10.60 & 6.94 & 12.16 & -12.16 & 6.11 \\
\ce{ CO2 } (Carbon dioxide) & 3.49 & -3.49 & 1.53 & 7.14 & -7.14 & 5.33 \\
\ce{ P2 } (Phosphorus diatomic) & 5.66 & -5.66 & 1.88 & 6.50 & -6.50 & 1.61 \\
\ce{ S2 } (Sulfur diatomic) ($^{3}{\Sigma}_{g}$) & 6.35 & -6.35 & 2.24 & 7.20 & -7.20 & 1.90 \\
\ce{ Cl2 } (Chlorine diatomic) & 6.59 & -6.59 & 3.12 & 7.70 & -7.70 & 2.45 \\
\ce{ NaCl } (Sodium Chloride) & 2.94 & -2.94 & 1.31 & 5.60 & -5.60 & 2.35 \\
\ce{ SiO } (Silicon monoxide) & 5.35 & -5.35 & 2.48 & 6.93 & -6.93 & 1.81 \\
\ce{ CS } (Carbon monosulfide) & 6.56 & -6.56 & 3.17 & 7.64 & -7.64 & 2.51 \\
\ce{ ClO } (Monochlorine monoxide) & 8.19 & -8.19 & 4.24 & 9.11 & -9.11 & 3.73 \\
\ce{ ClF } (Chlorine monofluoride) & 7.58 & -7.58 & 4.10 & 8.99 & -8.99 & 3.50 \\
\ce{ Si2H6 } (Disilane) & 2.56 & -2.56 & 0.98 & 4.69 & -4.69 & 2.08 \\
\ce{ CH3Cl } (Methyl chloride) & 2.91 & -2.91 & 3.01 & 5.81 & -5.81 & 2.96 \\
\ce{ CH3SH } (Methanethiol) & 2.76 & -2.76 & 1.58 & 5.25 & -5.25 & 1.48 \\
\ce{ SO2 } (Sulfur dioxide) & 7.36 & -7.36 & 2.66 & 8.27 & -8.27 & 2.17 \\
\ce{ BF3 } (Borane, trifluoro-) & 3.15 & -3.15 & 1.10 & 8.06 & -8.06 & 3.35 \\
\ce{ BCl3 } (Borane, trichloro-) & 5.95 & -5.95 & 2.44 & 6.63 & -6.63 & 3.12 \\
\ce{ AlCl3 } (Aluminum trichloride) & 4.01 & -4.01 & 1.89 & 5.55 & -5.55 & 1.46 \\
\ce{ CF4 } (Carbon tetrafluoride) & 2.91 & -2.91 & 0.59 & 5.52 & -5.52 & 0.23 \\
\ce{ CCl4 } (Carbon tetrachloride) & 4.96 & -4.96 & 2.04 & 5.68 & -5.68 & 1.70 \\
\ce{ OCS } (Carbonyl sulfide) & 4.59 & -4.59 & 4.95 & 7.49 & -7.49 & 2.48 \\
\ce{ CS2 } (Carbon disulfide) & 5.56 & -5.56 & 2.56 & 6.47 & -6.47 & 2.62 \\
\ce{ CF2O } (Carbonic difluoride) & 4.83 & -4.83 & 7.22 & 9.40 & -9.40 & 3.35 \\
\ce{ SiF4 } (Silicon tetrafluoride) & 3.64 & -3.64 & 1.53 & 6.16 & -6.16 & 1.35 \\
\ce{ N2O } (Nitrous oxide) & 5.08 & -5.08 & 6.59 & 8.93 & -8.93 & 3.25 \\
\ce{ NF3 } (Nitrogen trifluoride) & 5.72 & -5.72 & 8.00 & 9.38 & -9.38 & 5.11 \\
\ce{ PF3 } (Phosphorus trifluoride) & 3.73 & -3.73 & 3.96 & 6.75 & -6.75 & 2.74 \\
\ce{ O3 } (Ozone) & 8.68 & -8.68 & 3.79 & 9.49 & -9.49 & 3.33 \\
\ce{ F2O } (Difluorine monoxide) & 9.12 & -9.12 & 5.63 & 10.40 & -10.40 & 4.64 \\
\ce{ ClF3 } (Chlorine trifluoride) & 7.67 & -7.67 & 3.79 & 8.70 & -8.70 & 3.27 \\
\ce{ C2F4 } (Tetrafluoroethylene) & 3.25 & -3.25 & 7.30 & 8.25 & -8.25 & 2.83 \\
\ce{ CH3CCH } (Propyne) & 2.11 & -2.11 & 0.97 & 5.06 & -5.06 & 3.61 \\
\ce{ CH2CCH2 } (Allene) & 2.86 & -2.86 & 5.26 & 6.34 & -6.34 & 2.86 \\
\ce{ C3H4 } (Cyclopropene) & 3.23 & -3.23 & 5.55 & 7.16 & -7.16 & 2.83 \\
\ce{ C3H6 } (Cyclopropane) & 2.24 & -2.24 & 0.70 & 4.62 & -4.62 & 0.61 \\
\ce{ CH2F2 } (Methane, difluoro-) & 2.43 & -2.43 & 0.89 & 5.18 & -5.18 & 3.60 \\
\ce{ CHF3 } (Methane, trifluoro-) & 2.71 & -2.71 & 1.03 & 5.34 & -5.34 & 1.51 \\
\ce{ CH2Cl2 } (Methylene chloride) & 3.67 & -3.67 & 3.75 & 6.11 & -6.11 & 2.36 \\
\ce{ CHCl3 } (Chloroform) & 4.44 & -4.44 & 2.92 & 5.94 & -5.94 & 1.73 \\
\ce{ CH3NO2 } (Methane, nitro-) & 5.57 & -5.57 & 5.92 & 8.08 & -8.08 & 3.73 \\
\ce{ CH3SiH3 } (Methyl silane) & 2.57 & -2.57 & 1.14 & 4.83 & -4.83 & 0.76 \\
\ce{ HCOOH } (Formic acid) & 3.90 & -3.90 & 7.07 & 8.40 & -8.40 & 3.50 \\
\ce{ CH3CONH2 } (Acetamide) & 2.61 & -2.61 & 3.76 & 6.01 & -6.01 & 4.46 \\
\ce{ C2N2 } (Cyanogen) & 6.84 & -6.84 & 2.06 & 7.54 & -7.54 & 1.70 \\
\ce{ CH2CO } (Ketene) & 4.39 & -4.39 & 6.21 & 7.90 & -7.90 & 3.38 \\
\ce{ C2H4O } (Ethylene oxide) & 2.27 & -2.27 & 0.79 & 4.76 & -4.76 & 1.16 \\
\ce{ C2H2O2 } (Ethanedial) & 6.83 & -6.83 & 2.52 & 7.59 & -7.59 & 2.21 \\
\ce{ CH3CH2OH } (Ethanol) & 2.44 & -2.44 & 1.29 & 4.97 & -4.97 & 1.70 \\
\ce{ CH3OCH3 } (Dimethyl ether) & 2.24 & -2.24 & 0.88 & 4.67 & -4.67 & 0.99 \\
\ce{ C2H4S } (Thiirane) & 3.07 & -3.07 & 4.23 & 5.99 & -5.99 & 2.41 \\
\ce{ CH2CHF } (Ethene, fluoro-) & 3.31 & -3.31 & 5.73 & 7.32 & -7.32 & 2.67 \\
\ce{ CH3CH2Cl } (Ethyl chloride) & 2.59 & -2.59 & 2.47 & 5.49 & -5.49 & 3.28 \\
\ce{ CH2CHCl } (Ethene, chloro-) & 3.58 & -3.58 & 5.10 & 6.83 & -6.83 & 2.64 \\
\ce{ CH3COCl } (Acetyl Chloride) & 4.57 & -4.57 & 5.83 & 7.31 & -7.31 & 3.36 \\
\ce{ NO2 } (Nitrogen dioxide) & 7.72 & -7.72 & 4.08 & 9.15 & -9.15 & 3.29 \\
\ce{ CFCl3 } (Trichloromonofluoromethane) & 5.05 & -5.05 & 2.73 & 6.15 & -6.15 & 2.06 \\
\ce{ CF3Cl } (Methane, chlorotrifluoro-) & 4.27 & -4.27 & 5.52 & 7.41 & -7.41 & 3.37 \\
\ce{ HCCF } (Fluoroacetylene) & 2.62 & -2.62 & 2.37 & 5.67 & -5.67 & 2.28 \\
\ce{ HCCCN } (Cyanoacetylene) & 5.86 & -5.86 & 2.26 & 6.94 & -6.94 & 1.63 \\
\ce{ C4N2 } (2-Butynedinitrile) & 5.84 & -5.84 & 1.47 & 6.34 & -6.34 & 1.24 \\
\ce{ C2N2 } (Cyanogen) & 6.84 & -6.84 & 2.06 & 7.54 & -7.54 & 1.70 \\
\ce{ C3O2 } (Carbon suboxide) & 6.75 & -6.75 & 1.95 & 7.35 & -7.35 & 1.71 \\
\ce{ FCN } (Cyanogen fluoride) & 4.10 & -4.10 & 2.84 & 6.83 & -6.83 & 2.43 \\
\ce{ C4H2 } (Diacetylene) & 5.17 & -5.17 & 2.23 & 6.40 & -6.40 & 1.58 \\
\ce{ H2CS } (Thioformaldehyde) & 6.40 & -6.40 & 2.82 & 7.39 & -7.39 & 2.40 \\
\ce{ CHONH2 } (Formamide) & 2.98 & -2.98 & 7.25 & 6.66 & -6.66 & 4.49 \\
\ce{ CH2CCl2 } (Ethene, 1,1-dichloro-) & 3.97 & -3.97 & 4.42 & 6.43 & -6.43 & 3.10 \\
\ce{ C2HF3 } (Trifluoroethylene) & 3.23 & -3.23 & 6.54 & 7.79 & -7.79 & 2.86 \\
\ce{ CH2CF2 } (Ethene, 1,1-difluoro-) & 3.10 & -3.10 & 6.19 & 7.46 & -7.46 & 2.87 \\
\ce{ CH3F } (Methyl fluoride) & 2.33 & -2.33 & 0.81 & 5.34 & -5.34 & 2.66 \\
\ce{ CF2Cl2 } (Difluorodichloromethane) & 4.80 & -4.80 & 3.72 & 6.69 & -6.69 & 2.52 \\
\ce{ SiF2 } (Silicon difluoride) & 5.53 & -5.53 & 2.55 & 7.24 & -7.24 & 1.91 \\
\hline
MSE &  & -4.89 & 3.11 &  & -7.23 &  2.70 \\ 
MAE &  & 4.89 &   3.11 &  & 7.23 &  2.70 \\
rms   &  & 5.36 &   3.73 &  & 7.48 &  3.01 \\ 	 	 	 	
\end{tabular}
\end{ruledtabular}
\end{table}

\end{document}